\documentclass[aps,prc,twocolumn,superscriptaddress,showpacs,showkeys]{revtex4-1}
\usepackage{graphicx}

\begin{document}
\title{Search for solar axions produced by Compton process and bremsstrahlung using axioelectric effect}
\author{A.V.~Derbin}
\author{I. S.~Drachnev}
\author{A.S.~Kayunov}
\author{V.N.~Muratova}
\affiliation{St.Petersburg Nuclear Physics Institute, Gatchina, Russia 188300}
\begin{abstract}

The axio-electric effect in silicon atoms is sought for solar axions appearing owing to bremsstrahlung and the Compton process. Axions are detected
using a Si(Li) detector placed in a low-background setup. As a result, a model-independent constraint on the axion-electron coupling constant
$|g_{Ae}| \leq 2.2\times 10^{-10}$ has been obtained, which leads to the bounds $m_{A}\leq$ 7.9 eV and $m_{A}\leq$ 1.4 keV for the mass of the axion
in the DFSZ and KSVZ models, respectively (90\% C.L.).

\end{abstract}


\maketitle

\section{ INTRODUCTION}
The axion is a neutral pseudoscalar particle introduced in the theory in order to explain the absence of  the $CP$ violation in strong interactions
\cite{Peccei:1977}- \cite{Wilczek:1978}. The effective axion-photon ($g_{A\gamma}$), axion-lepton ($g_{Ae}$), and axion-hadron ($g_{AN}$) coupling
constants, as well as the mass of the axion $m_{A}$, are inversely proportional to the energy $f_{A}$ at which $U_{PQ}$(1) symmetry introduced in
\cite{Peccei:1977} is broken. The initial model of the standard PQWW -axion, where $f_{A} =(\sqrt{2}G_F)^{-1/2}$ is fixed at the electroweak scale,
was reliably excluded by reactor and accelerator experiments \cite{Nakamura:2010}.

    New models of an invisible axion with an arbitrary scale of $U_{PQ}$ symmetry breaking stimulated continuation of the
experimental search for a pseudo-scalar particle with a mass in a wide range from $10^{-16}$ to $10^{6}$ eV. The models of the
KSVZ axion (hadron axion) \cite{Kim:1979, Shifman:1980} and the DFSZ axion (GUT axion) \cite{Zhitninskii:1980, Dine:1981} are the
two main models of the invisible axion. The mass of the axion $m_{A}$ in both models is expressed in terms of the properties of
the ${\pi}^0$ meson as
\begin{equation}
  m_{A}=\frac{f_{\pi}m_{\pi}}{f_{A}}\left[\frac{z}{(1+z+w)(1+z)}\right]^{1/2},
\end{equation}
where $m_{\pi}$ and $f_{\pi}$ are the mass and decay constant of the $\pi^0$ meson, respectively, and $z = m_{u}/m_{d}\approx
0.56$ and $w = m_{u}/m_{s} \approx 0.029$ are the quark mass ratios.

The most stringent bounds on the interaction of the axion with matter were obtained from astrophysical and cosmological data.
Astrophysical constraints based on the interaction of axions with photons and electrons lead to the bound $m_{A} < 10^{-2}$ eV if
the standard relation between $g_{Ae}$, $g_{A\gamma}$, and $m_{A}$  is used\cite{Nakamura:2010, Raffelt:2006}. The SN1987A data
 provide upper and lower bounds for the axion-nucleon coupling constant\cite{Raffelt:1999}. These bounds allow the axion mass
range $\approx 10$ eV (the so-called hadron axion window). At the same time, the latest cosmological data imply that $m_{A} < 1$
eV \cite{Raffelt:2001}.

Experimental bounds on the mass of the axion follow from constraints on the $g_{A\gamma}$, $g_{Ae}$, and $g_{AN}$ coupling
constants, which significantly depend on the theoretical model used. In particular, the hadron axion does not interact with
leptons and usual quarks at the tree level. This leads to the strong suppression of the interaction of the axion with leptons,
which is due only to radiative corrections. In some models, the axion-photon coupling constant $g_{A\gamma}$ is strongly
different from the initial values for the models of the DFSZ and KVSZ axions \cite{Kaplan:1980}.

This work is devoted to the search for axions appearing in the Sun owing to the axion-electron interaction, namely, bremsstrahlung from electrons in
the field of a nucleus, $e^{-} + Z \rightarrow Z + A$, and the Compton process, $\gamma + e^{-}\rightarrow e^{-} + A$
\cite{Kato:1975}-\cite{Kekez:2009}. The reaction cross sections and, therefore, axion fluxes are proportional to $g^{2}_{Ae}$.

To detect axions, we used the axio-electric absorption of an axion by atoms, which is an analog of the photoelectric effect. Since the reaction cross
sections are proportional to $g^{2}_{Ae}$, the expected count rate of axions depends only on the axion-electron coupling constant and is proportional
to $g^{4}_{Ae}$. The dimensionless constant $g_{Ae}$ is generally related to the electron mass and $f_{A}$ as $g_{Ae}$ = $C_{e}m_{e}/f_{A}$, where
$C_{e}$ is the model -dependent parameter of about 1. In the DFSZ axion model, $g_{Ae}$ is related to the electron mass $m_e$ as
 \begin{equation}
g_{Ae}=(1/3)(cos^{2}\beta)m_{e}/ f_{A},
\end{equation}
where $\beta$ is an arbitrary angle. In the KSVZ axion model, $g_{Ae}$ is much smaller (by a factor of about $\alpha^{2}$),
because it is determined only by radiative corrections \cite{Srednicki:1985}:
\begin{equation}
g_{Ae}=\frac{3\alpha^{2}nm_e}{2\pi
f_{A}}\left(\frac{E}{N}\ln\frac{f_{A}}{m_e}-\frac{2}{3}\frac{4+z+w}{1+z+w}\ln\frac{\Lambda}{m_e}\right),\label{Gaee}
\end{equation}
where $n$ is the number of generations, E/N is the model dependent ratio of the order of unity and $\Lambda \approx 1$ GeV is the QCD cutoff scale.

\section{ SPECTRA OF SOLAR AXIONS AND THE CROSS SECTION FOR THE AXIO-ELECTRIC EFFECT}
The spectrum of Compton axions ($d\Phi^{Comp}_{A} /dE_{A}$) appearing in the $\gamma + e^{-}\rightarrow e^{-} + A$ reaction was calculated by
integrating the Compton conversion cross section obtained in \cite{Pospelov:2008, Gondolo:2009} over the Planck energy distribution of thermal
photons taking into account the radial distribution of the temperature and electron density in the standard solar model BSB05 \cite{Bahcall:2005} for
the high-metallicity case \cite{Asplund:2006}. The calculation results for various masses of the axions are shown in Fig.\ref{Figure:Compton
process}.
\begin{figure}
\includegraphics[width=9cm,height=10.5cm]{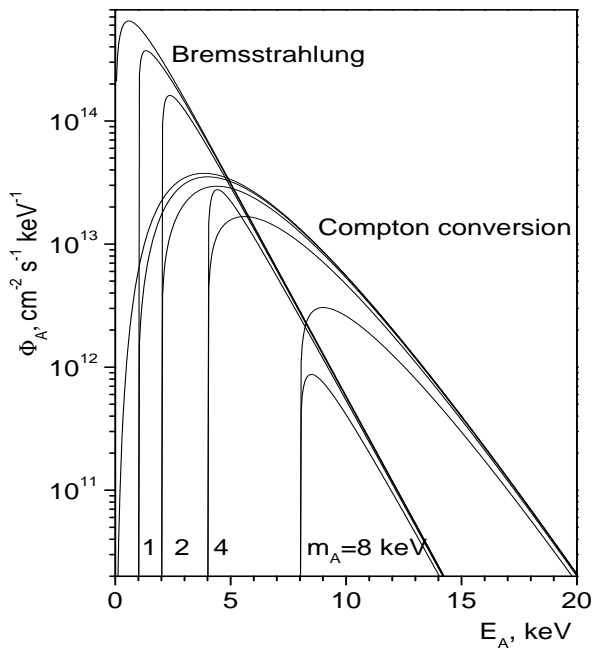}
\caption {Spectra of solar axions produced by the bremsstrahlung and the Compton process as calculated for various values of the mass of the axion
and constant $g_{Ae} = 10^{-10}$.} \label{Figure:Compton process}
\end{figure}

The spectrum of bremsstrahlung axions ($d\Phi^{Brem}_{A} /dE_{A}$) was calculated by the same method. The cross section for the $e^{-} + Z\rightarrow
Z + A$ reaction was obtained in \cite{Zhitnitsky:1979}. The Maxwell-Boltzmann distribution was used to describe the energy spectrum of electrons.
Bremsstrahlung was considered in the field of H, $\rm^{4}He$, $\rm^{3}He$, $\rm^{12}C$, $\rm^{14}N$, and $\rm^{16}O$ nuclei. The spectrum of
bremsstrahlung axions is softer (Fig.\ref{Figure:Compton process}).

The calculation procedures were described in detail in \cite{Derbin:2011}, where simple parameterizations describing the spectra with an accuracy of
1\% in the range of (1--10) keV were presented. The spectra of axions shown in Fig. 1 were calculated for $g_{Ae} = 10^{-10}$. The average energies
of axions are 1.6 keV and 5.1 keV for bremsstrahlung and Compton axions, respectively, for the case $m_{A} \approx 0$.   The maximum intensity is
reached at energies of 0.6 keV and 3.8 keV, respectively. The axion flux almost vanishes at energies above 20 keV.

\begin{figure}
\includegraphics[width=9cm,height=10.5cm]{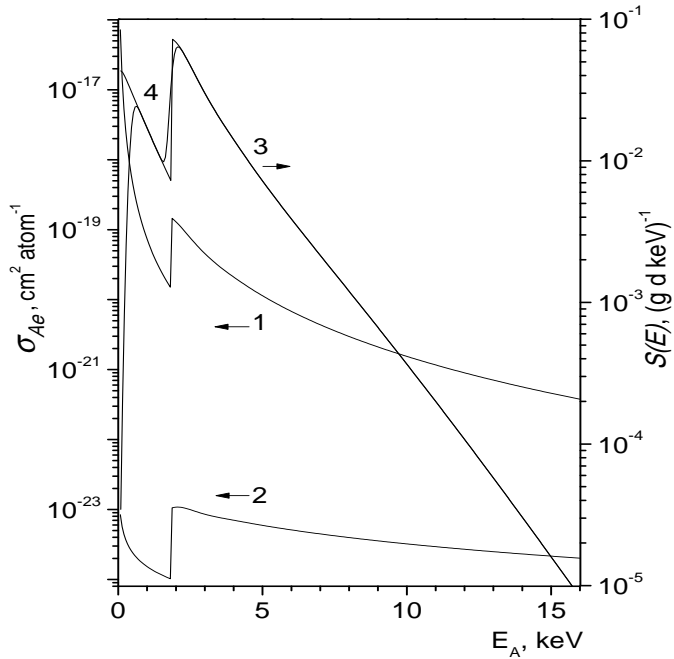}
\caption {Cross sections for the (1) photoelectric effect for a silicon atom and (2) axioelectric effect for $g_{Ae}=1$ and $m_{A}=0$, (3) the
expected spectrum of events detected per day in 1 g of Si in 1 keV interval, and (4) the spectrum taking into account the resolution of the detector
(right scale).} \label{Figure:Cross sections}
\end{figure}

An axion interacting with an electron should undergo axio-electric absorption, which is an analog of the photoelectric effect. Silicon atoms entering
into the composition of a Si(Li) detector were used in our experiment as targets for the axio-electric effect. The cross section for the
axio-electric effect for nonrelativistic axions is proportional to the cross section for the photoelectric effect $\sigma_{pe}$ for photons with the
energy equal to the mass of the axion \cite{Pospelov:2008}:
\begin{equation}
\sigma_{ae}(E_{A})=\sigma_{pe}(E_{\gamma} = m_{A})\frac{g^{2}_{Ae}}{\beta}\frac{3m^{2}_{A}}{16\pi \alpha m^{2}_{e}}
\end{equation}
  where $\beta= v/c = p_{A}/E_{A}$ is the
velocity of the axion. For relativistic axions in the case $E_{A} < m_{e}$ and $m_{A}\rightarrow 0$, the cross section differs from Eq. (4) by a
factor of about $2/3$ and by a change of $m_{A}$ to $E_{A}$ \cite{Pospelov:2008, Derevianko:2010}. The cross section for the photoelectric effect
$\sigma_{pe}$(E) for various energies of photons was calculated using the XCOM database \cite{Berger:1}. Since the efficiency of the detection of an
appearing electron and accompanying X rays by a Si(Li) detector is $\cong 100\%$ the expected spectrum of the detected energy is given by the
expression
\begin{equation}
\frac{dN}{dE}=\sigma_{ae}(E_{A},m_{A})\left(\frac{d\Phi^{Brem}_{A}}{dE_{A}}+\frac{d\Phi^{Comp}_{A}}{dE_{A}}\right)
\end{equation}
To compare with the experimentally measured spectrum, Eq. (5) should be additionally averaged using the response function of the Si(Li) detector,
which is well described by a Gaussian for photons and electrons in the energy range of (0--20) keV. The cross section for the axio-electric effect
calculated for $m_{A} = 0$ is shown in Fig.\ref{Figure:Cross sections} along with the expected spectrum from axions $S(E,m_{A})$ with and without the
inclusion of the energy resolution of the Si(Li) detector ($\sigma = 115 ~\rm{eV}$). The spectrum has a characteristic feature at an energy of 1.84
keV, which corresponds to the binding energy of electrons in the K shell. The peak shape of the feature significantly increases the sensitivity of
the experiment compared to the case of a smooth curve.

The number of detected electrons and X-ray photons following the absorption of an axion depends on the number of silicon atoms in the detector, the
time of measurements, and the efficiency of the detector. The probability of observing a signal is determined by the background of the experimental
setup.

\section{  EXPERIMENTAL SETUP}
In the experiment, we used a Si(Li) detector with a sensitive-region diameter of 17 mm and a thickness of 2.5 mm. The detector was placed in a vacuum
cryostat with the input beryllium window 20 $\mu$m thick. The window was used for energy calibration and determination of the detection efficiency of
gamma-ray photons in order to find the sensitive volume of the detector.

The detector was surrounded by 12.5 cm of copper and 2.5 cm of lead, which reduced the background of the detector at an energy of 14 keV by a factor
of 110 as compared to the unshielded detector. In order to suppress the background from cosmic rays and fast neutrons, we used five scintillation
detectors, which closed the detector almost completely except for the bottom side, where a Dewar vessel with liquid nitrogen was placed.

The spectrometric channel was organized as follows. A field-effect transistor was mounted on a Teflon plate near the Si(Li) detector and was cooled
to a temperature close to liquid nitrogen. A preamplifier except for the first cascade was placed outside the passive shielding in order to reduce
the background. A signal after the preamplifier was guided to two amplifiers with different amplification coefficients, which made it possible to
measure the spectrum of signals in the low energy (0.5--60) keV and harder (10--500) keV regions. The measurement of two regions allowed the reliable
control of the background in the natural radioactivity region. Each spectroscopic channel was equipped with its own analog-to-digital converter.
Thus, four 4096-channel spectra were simultaneously accumulated in computer memory (two energy ranges in coincidence and anticoincidence with an
active shielding signal).

\begin{figure}
\includegraphics[width=9cm,height=10.5cm]{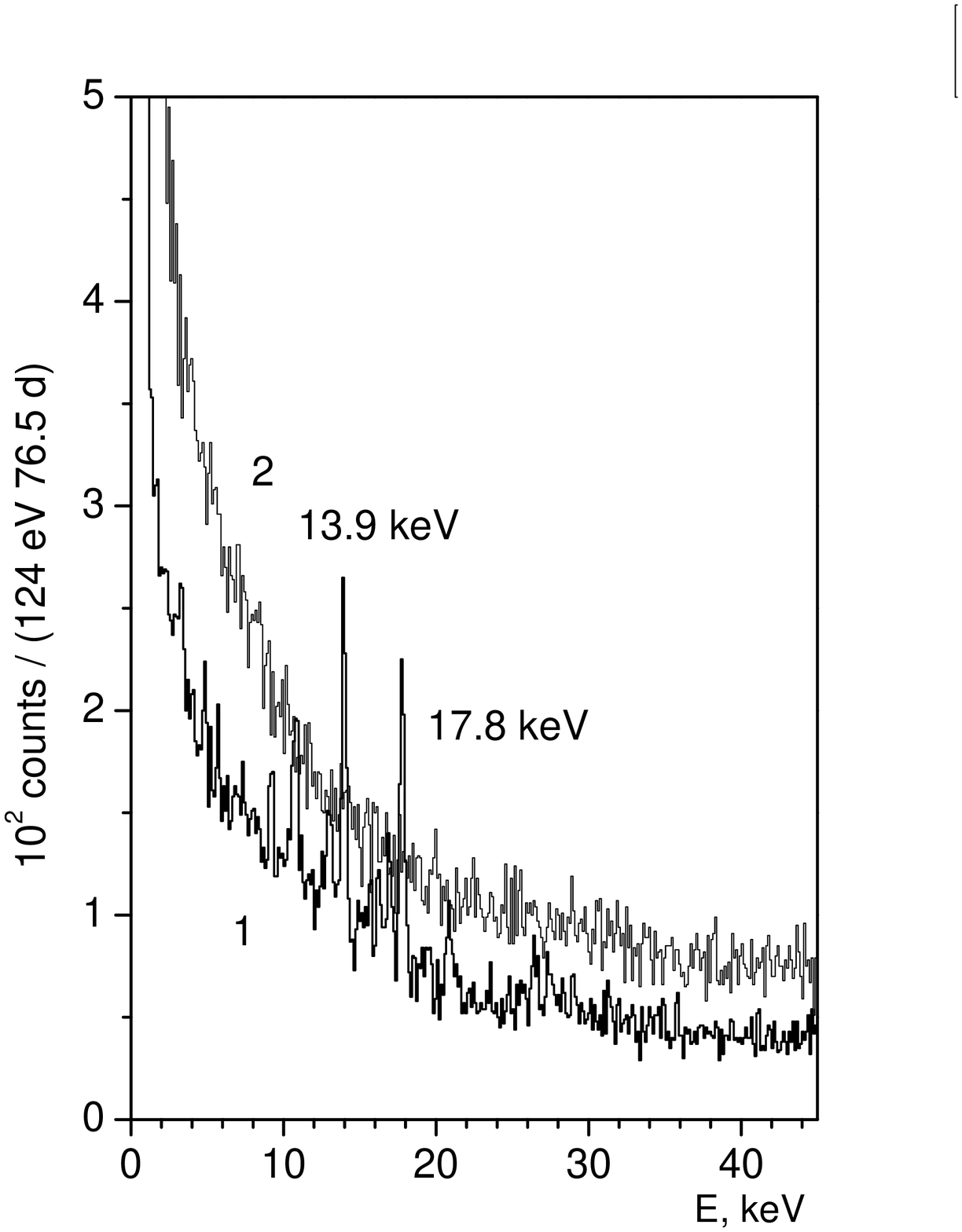}
\caption {Spectra of the Si(Li) detector measured in (1) anticoincidence and (2) coincidence with the active shielding signal.}
\label{Figure:Spectra SiLi}
\end{figure}

\begin{figure}
\includegraphics[width=9cm,height=10.5cm]{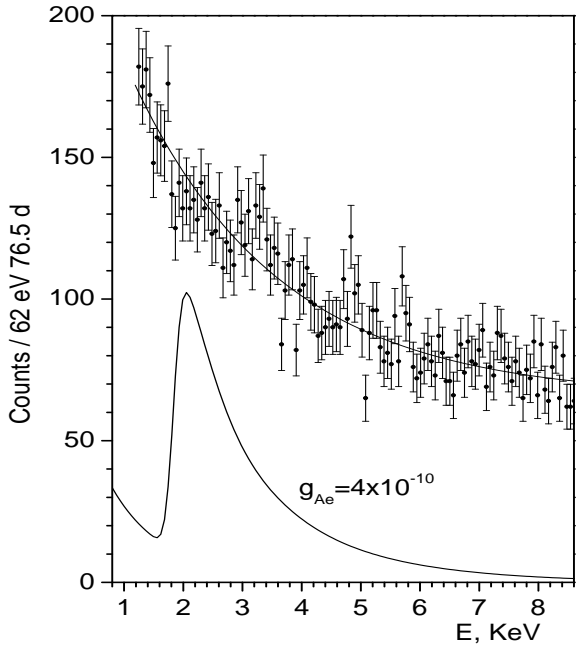}
\caption {Spectrum of the signals of the Si(Li) detector with the optimal fit and (solid line) the expected spectrum in the case
of the detection of axions with $m_{A} \approx 0$ and $g_{Ae} = 4 \times 10^{-10}$.} \label{Figure:Expected spectrum}
\end{figure}
The sensitive volume of the detector was determined using $\gamma$ lines of standard $\rm^{57}Co$ and $\rm^{241}Am$ sources. The
determined number of silicon atoms in a sensitive volume of $0.57 \pm 0.03$ $\rm cm^{3}$ was $\rm N_{Si} =2.85\times 10^{22}$.
The error is mainly due to the inaccuracy of the position of the $\rm^{57}Co$ source with respect to the detector.

\section{ RESULTS}

Measurements continued for 76.5 days of pure time in the form of two-hour runs in order to control the stability of the operation of the
spectrometric channels of the Si(Li) detector and active shielding scintillation detectors. Figure \ref{Figure:Spectra SiLi} shows the total energy
spectrum of signals in the range of (0.8--45) keV. The spectrum of events detected in anticoincidence with active shielding signals clearly  exhibits
the 13.9 and 17.8 keV Np X-ray lines attributed to decays of $\rm^{241}Am$. The spectrum of events detected in coincidence with active shielding
signals has no statistically significant peaks.

The spectrum measured in the range of (0.8--9) keV is shown in Fig.\ref{Figure:Expected spectrum} along with the expected spectrum $S(E, m_{A})$ in
the case of the detection of theaxioelectric effect for $m_{A}\approx 0$ and $g_{Ae} = 4\times 10^{-10}$ (line 4 in Fig.\ref{Figure:Cross sections}).
It can be seen that the shown characteristic feature is not observed in the measured spectrum. The maximum likelihood method is used to determine the
intensity of the response function.
\begin{figure}
\includegraphics[width=9cm,height=10.5cm]{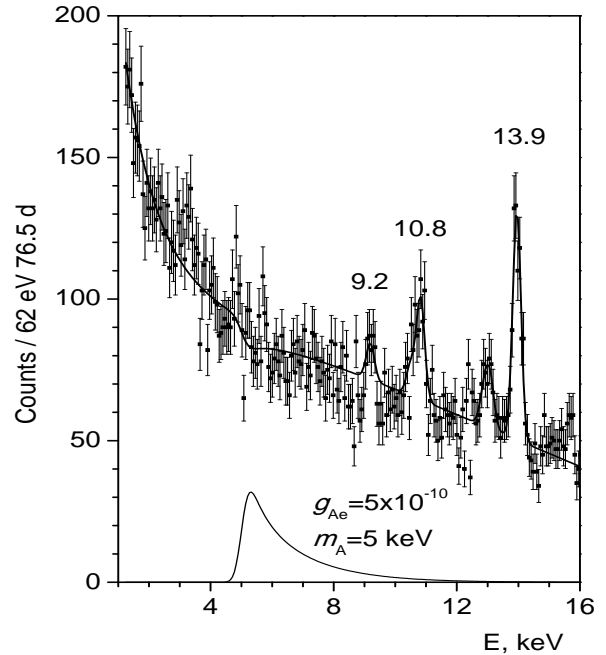}
\caption {Spectrum of the signals of the Si(Li) detector in the range of (1--16) keV with the optimal fit for $m_{A} = 5$ keV. The energies of the
Gaussian peaks are given near them. The expected spectrum is shown for the case of the detection of axions with $m_{A} = 5$ keV and $g_{Ae} = 5
\times 10^{-10}$.} \label{Figure:Fit_for 5 keV}
\end{figure}
\begin{figure}
\includegraphics[width=9cm,height=10.5cm]{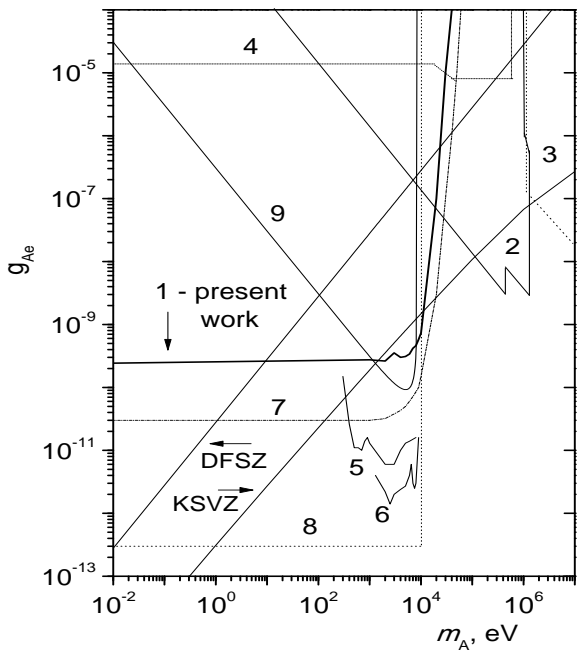}
\caption {Bounds for the axion-electron coupling constant from (1) this work, (2) reactor experiments and solar axions with energies of 0.478 and 5.5
MeV \cite{Altmann:1995}-\cite{Bellini:2012}, (3) beam dump experiments \cite{Konaka:1986,Bjorken:1988}, (4) decay of orthopositronium
\cite{Asai:1991}, (5) CoGeNT \cite{Aalseth:2008}, (6) CDMS \cite{Ahmed:2009}, (7) bound for the axion luminosity of the Sun \cite{Gondolo:2009}, (8)
red giants \cite{Raffelt:2008}, and (9) experiment with $\rm ^{169}Tm$ \cite{Derbin:2011}. The regions of excluded values lie above the corresponding
lines. The inclined lines show the $g_{Ae}$ values in the DFSZ and KSVZ ($E/N=8/3$) models.}
 \label{Figure:Bounds}
\end{figure}
The experimental spectrum was represented in the form of the sum of an exponential describing the smooth background and the response function for
axions $S(E, m_{A})$:
\begin{equation}
N(E) = a+b\exp(cE)+g^{4}_{Ae}S(E,m_{A})N_{Si}T
\end{equation}
Here, $N_{Si}$ is the number of silicon atoms in the sensitive volume of the detector and $T = 6.61 \times 10^{6}$ is the live time of measurement.

The energy scale and energy resolution $\sigma$ were determined from peaks manifested  in the measurements and were fixed. The constant $g^{4}_{Ae}$
and the parameters $a$, $b$, and $c$ that describe the continuous background were free parameters. The fit in the range of (1.2--8.0) keV for $m_{A}
= 0$ is shown in Fig.\ref{Figure:Expected spectrum}. The minimum value $\chi^2 = 111/108$ corresponds to the value $g^{4}_{Ae}$ = $(-0.06 \pm 15)
\times 10^{40}$. To determine the upper bound for $g^{4}_{Ae}$, we used the standard method: the $\chi^{2}$ value was determined by fixed
$g^{4}_{Ae}$ values, whereas the other parameters were free. The resulting probability function $P(\chi^{2}(g^{4}_{Ae} ))$ was normalized to unity
for values $g^{4}_{Ae}\geq 0$. The upper bound for $|g_{Ae}|$ is
\begin{equation}
|g_{Ae}|\leq 2.2 \times  10^{-10}\label{mainlimit}
\end{equation}
at 90\% C.L. Limit (7) is a model independent bound for the coupling constant of the axion or any other pseudoscalar relativistic particle with the
electron.

If the mass of the axion is several keV, the expected spectrum in the detector changes significantly and depends on the
particular $m_{A}$ value. To determine values, we used the procedure similar to that described above. The spectra of solar axions
were calculated for different values of $m_{A}$ from 1 to 10 keV with a step of 1 keV. The cross section of the axioelectric
absorption was calculated by the formula
\begin{equation}
\sigma_{abs}(E_{A})=\sigma_{pe}(E_{A})\frac{g^{2}_{Ae}}{\beta}\frac{3E^{2}_{A}}{16\pi\alpha
m_e^2}\left(1-\frac{\beta}{3}\right)
\end{equation}
At $\beta \rightarrow $ 1 and $\beta \rightarrow $ 0, this formula coincides with the cross sections for relativistic and
nonrelativistic axions obtained in [27] (Eq. (4)) and provides an extrapolation approximately linear in $\beta $, which ensures a
sufficient accuracy for the case under consideration.

An example of the response function of the detector for detecting axions is shown in Fig.\ref{Figure:Fit_for 5 keV}. The spectrum fitting range was
expanded to 16 keV. To describe the experimental spectrum in a wide range, function (6) was supplemented by a linear term for describing the
continuous background and six Gaussians for describing the peaks of the characteristic Np X rays manifested in the measurements
\cite{Derbin:2009,Derbin:2009A}. The fit for $m_{A}$ = 5 keV is shown in Fig.\ref{Figure:Fit_for 5 keV}.

The maximum deviation of $g^{4}_{Ae}$ from zero for all $m_{A}$ values is 2.5$\sigma$. The upper bounds obtained for $|g_{Ae}|$ at various $m_{A}$
values are shown in Fig.\ref{Figure:Bounds} (line 1) in comparison with the other experimental results.

For large masses of the axion $m_{A} > 10$ keV, the most stringent bounds were obtained from reactor \cite{Altmann:1995, Chang:2007} and accelerator
\cite{Konaka:1986, Bjorken:1988} experiments, as well as for high energy solar axions \cite{Bellini:2008,Derbin:2010,Bellini:2012}. For small $m_A$
values ($m_A < 0.1$) keV, the most stringent bounds were obtained from astrophysical data, namely, from the cooling rate  of red giants
\cite{Raffelt:2008} and under the assumption that the axion luminosity of the Sun does not exceed 10\% of the observed neutrino luminosity
\cite{Gondolo:2009}. The search for axions with masses in the intermediate region $m_{A}\approx 1$ keV was specially performed in view of the
possibility of attributing the DAMA experiment result to the axio-electric effect with silicon and germanium detectors designed for seeking dark
matter \cite{Aalseth:2008,Ahmed:2009}. Solar axions having a continuous energy spectrum should lead to the resonant excitation of low-lying nuclear
levels \cite{Derbin:2011}.

The upper limit (\ref{mainlimit}) obtained for $|g_{Ae}|$ lead to bounds for the mass of the axion that depend on a particular model. In the model of
the DFSZ axion, the relation between $g_{Ae}$ and $m_{A}$ for $cos^{2}\beta = 1$ has the form $g_{Ae}= 2.8 \times 10^{-11}m_{A}$, where $m_{A}$ is
expressed in eV units. The limit (7) leads to the bound $m^{DFSZ}_{A}\leq 7.9$ eV (at 90\% C.L.). The upper bound for the excluded $m_{A}$ values
units is 15 keV (Fig.\ref{Figure:Bounds}). In the model of the KSVZ axion, $g_{Ae}$ and $m_{A}$ are related through Eq. (3). For the ratio $E/N =
8/3$, which is characteristic of most variants of the DFSZ axion, and $n = 3$, the region of excluded values $m^{KSVZ}_{A}$ is (1.4--12) keV
(Fig.\ref{Figure:Bounds}).

\section{ CONCLUSION}

 The axio-electric absorption of solar axions, which produced by the Compton process and the bremsstrahlung, by silicon atoms has been sought.
 To detect the axio-electric effect, a Si(Li) detector placed in a low-background setup was used. As a result, new limits have been obtained
 for the axion-electron coupling constant and mass of the axion. For axions with a mass smaller than 1 keV, the resulting limit
 is $|g_{Ae}| \leq 2.2\times 10^{-10}$ at 90\% C.L..

\end{document}